\documentclass[preprint,review, 12pt]{elsarticle}

\usepackage{lineno,hyperref}

\usepackage{setspace}  

\journal{Materials Science and Engineering: B}

\usepackage{subfig}
\usepackage{soul}
\usepackage{epstopdf}
\begin{document}

\begin{frontmatter}

\title{Aluminium as catalyst for ZnO nanostructures growth}

\author{C. Zandalazini}
\address{Laboratorio de F\'{i}sica del S\'{o}lido, Departamento de
F\'{i}sica, Facultad de Ciencias Exactas y Tecnolog\'{i}a, Universidad Nacional de Tucum\'{a}n and
Consejo Nacional de Investigaciones Cient\'{i}ficas y T\'{e}cnicas, Argentina.}
\cortext[C. Zandalazini]{Corresponding author}
\ead{zc@famaf.unc.edu.ar}

\author{M. Villafuerte}
\address{Laboratorio de F\'{i}sica del S\'{o}lido, Departamento de
F\'{i}sica, Facultad de Ciencias Exactas y Tecnolog\'{i}a, Universidad Nacional de Tucum\'{a}n and
Consejo Nacional de Investigaciones Cient\'{i}ficas y T\'{e}cnicas, Argentina.}

\author{M. Oliva}
\address{Grupo de Ciencia de Materiales, Facultad de Matem\'{a}tica, Astronom\'{i}a, y F\'{i}sica, Universidad Nacional de C\'ordoba and Consejo Nacional de Investigaciones Cient\'{i}ficas y T\'{e}cnicas, Argentina.}

\author{S. P. Heluani}
\address{Laboratorio de F\'{i}sica del S\'{o}lido, Departamento de
F\'{i}sica, Facultad de Ciencias Exactas y Tecnolog\'{i}a, Universidad Nacional de Tucum\'{a}n, Argentina.}

\begin{abstract}
We report the growth of Al-catalysed ZnO nanowires (NWs) using a thermal evaporation technique.Before the growth, the substrates were covered with a distribution of Al nano-island that act as seeds. We found that the density of NWs increases as the density of seeds is increased, confirming the catalyst properties of Al. The critical parameters of growth are the substrate temperature, oxygen partial pressure and the thickness of the initial Al layer from which the seeds are formed. The results showed that the oxygen pressure has a strong influence on the structural characteristics: the nanowires exhibit a preferential orientation in the (00l)-planes when they are grown at low oxygen flow, and they become polycrystalline when a high concentration of oxygen in the flow is used. We consider that the growth occurs via a vapor-solid-solid (VSS) process as the predominant growth mechanism.
\end{abstract}


\end{frontmatter}

\section{Introduction}

During the last decade, there have been a renewed interest in studying zinc oxide since it is one of the most promising materials to be used in nanotechnology due to its exceptional photonic and electronic properties, as well as its great thermal stability and oxidation resistance \cite{Wang04,Osgur09}. The ability of the ZnO surface to adsorb gas molecules determines its application in gas sensor devices. Also, the lack of a center of symmetry in its wurtzite structure, combined with a large electromechanical coupling, results in strong piezoelectric and pyroelectric properties and the consequent use of ZnO in mechanical actuators and piezoelectric sensors. Undoubtedly, one of the most important characteristic is its wide band-gap (3.37 eV) which makes ZnO suitable for short wavelength optoelectronic applications.

There are several techniques to grow nanostructures (NSs). Considering the growth from the vapor phase, the vapor-liquid-solid (VLS) mechanism \cite{Wagn64,Klaus11} is one of the most used for growing nanostructures. In this technique, the anisotropic crystal growth is guided by a droplet of liquid metal, which acts as a preferred absorption site for the incoming vapor reactants due to its higher surface sticking coefficient.  Typically each metal droplet, which works as a virtual template, remains at the tip of the resultant nanostructure. In most cases, the diameter of the nanostructure is largely determined by the size of the metal droplet, and the latter depends primarily on temperature, degree of supersaturation, and on the liquid$-$vapor surface energy \cite{Klaus11}. 

On the other hand, the growth of NSs from the reactants in the gas phase could also occur in the absence of any metal catalyst, via the process known as vapor-solid (VS) mechanism \cite{Klaus11,Ramdir10,YWang06,Sears53}. In this case the morphology of the resultant NSs largely depends on the substrate temperature, pressure, carrier gas flow rate, and source materials \cite{Blakely62,Dai03}, and it has also been found that the NSs morphology is strongly determined by the anisotropy in the growth rates of the different crystallographic surfaces \cite{Klaus11}.

In addition to the VLS and VS, a third and less studied growth mechanism is also possible, namely the vapor-solid-solid process (VSS) \cite{YWang06,Campos08,Lensch09,Potie11}. Although the precise physical mechanism that governs the VSS growing process is still a controversial subject, it is accepted that the growth occurs through or at the surface of a solid catalyst. The VSS growth has been reported for many catalyst/nanostructure pairs, such as TiSi$_{2}$-catalysed Si nanowires \cite{Kamins00}, and SiO$_{x}$-catalysed InAS nanowires \cite{Mandl06}, among others. Unlike the VLS mechanism, the VSS process occurs at temperatures lower than the bulk eutectic temperature. In particular, Y. Wang \textit{et al.} \cite{YWang06} have reported that Al can be used as a catalyst to grow Si NWs at temperatures considerably lower than the eutectic temperature of the Al-Si binary system (577$^\circ$C). Unlike the typical VLS growth, Al-catalysed NWs tend to be tapered, a feature commonly associated with the VSS growth mechanism \cite{Potie11}. 

Recently, K. Govatsi \textit{et al.} \cite{Govatsi14} have published a detailed study on the influence of the Au film thickness and annealing conditions on the VLS growth of ZnO NSs. This work can be seen as an indication of both the considerable current interest in growing ZnO NSs and of the difficulties in the reproducibility of the results, even when gold, one of the most studied catalyst, is used. This in turn is due to the intrinsic difficulties in controlling the many parameters that rule the process. \\
In this article, we study the growth of ZnO nanowires using aluminium as catalyst as an alternative to the conventional ones. From a technological point of view, aluminium is a much more attractive catalyst material, and even though it is known that aluminium quickly oxidizes, our results show that this is not an impediment to act as a solid catalyst. We study the correlation between the density of Al seeds and the density of the resulting NWs. Also, in order to investigate the growth mode, we studied the dependence of the form, density and structure of the NWs on the growing conditions using different temperatures and partial pressures of oxygen.

\section{Experimental Details}
\label{ed}

Films of high purity aluminium (5N) were prepared at room temperature using DC sputtering deposition on SiO$_{2}$(500nm)/(100)Si substrates. The substrates were previously cleaned by a three-step ultrasonic cleaning, using acetone, ethanol and deionized water in each three-minute step. After deposition, the films were annealed at 600$^{\circ}$C in argon atmosphere (760 Torr) for 15 minutes, in order to obtain the nano-islands (seeds). 

\begin{figure} 
\includegraphics[width=1\columnwidth]{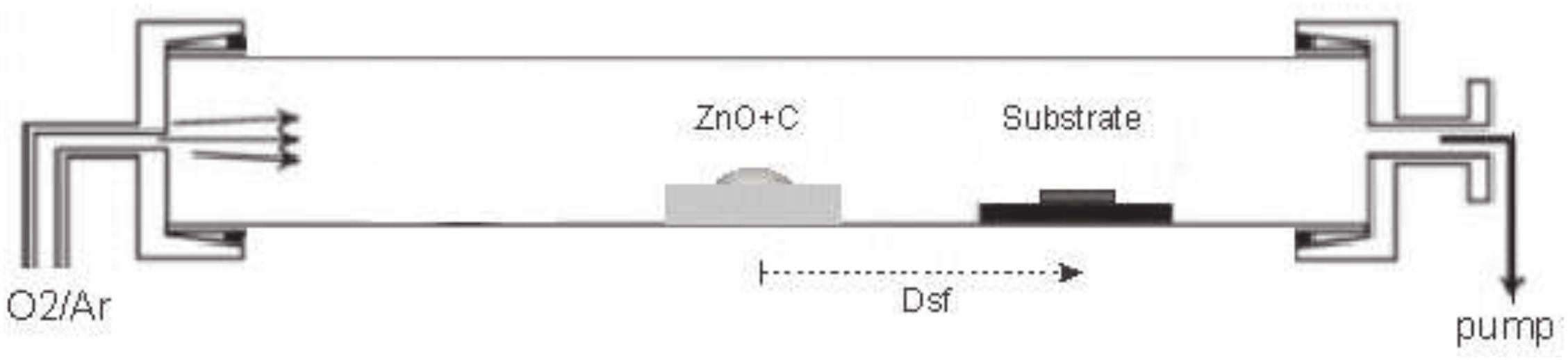}
\caption{Schematic diagram of the ZnO NSs growing system. Distance $D_{sf}$ determines the substrate temperature for a given central temperature.} \label{vls}
\end{figure}

A mixture (1:1 weight ratio) of ZnO powder (Alfa-Aesa ZnO 5N) and graphite C (impurity $<$ 0.1 ppm, Ringsdorff GmbH) was placed in an alumina boat in the center of the quartz tube surrounded by a tube furnace, which was maintained at 1100$^{\circ}$C. Through one of the quartz tube ends, different oxygen/argon high purity (5N) gas mixtures were injected using a total flow of 150 sccm. A pressure of 1 Torr was maintained during the growth. The seeded substrates were placed downstream on top of an upside-down alumina boat, at temperatures between 800 and 950 $^{\circ}$C, controlled by the distance $D_{sf}$ to the center of the furnace, see Fig. \ref{vls}.




The morphology and chemical analysis of the samples and substrates were studied by field-emission scanning electron microscopy (FEG-SEM, ZEISS Supra 55VP with Energy-dispersive X-ray spectroscopy). For structural characterizations, X-ray diffraction (XRD) patterns were recorded in a Philips PW 3830 diffractometer using Cu-K$_{\alpha}$ radiation.

\section{Results and discussion}

\label{res}

\subsection{Deposition of the Al initial films and seeds formation}

\begin{figure}
\includegraphics[width=1\columnwidth]{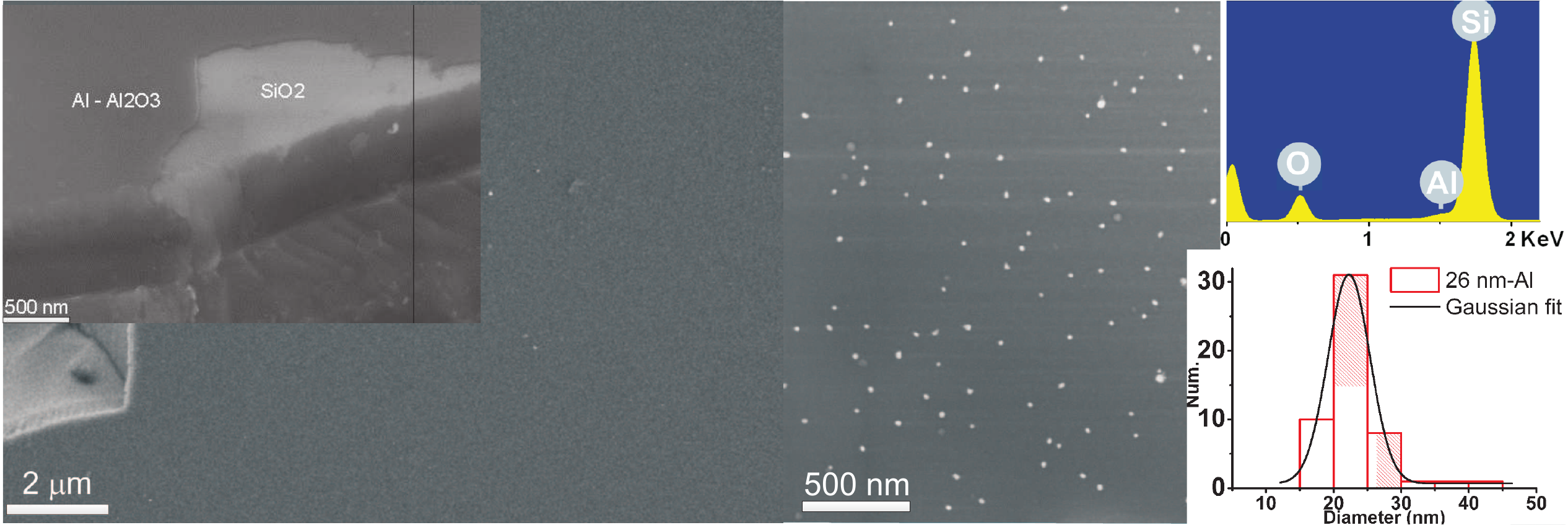}
\caption{Left: SEM image of the SiO$_{2}$/Si substrate covered with an Al film 26 nm thick. The inset shows a magnification of a zone where the SiO$_{2}$ is visible. Right: (26 nm)Al/SiO$_{2}$/Si after an ex-situ annealing showing the resulting droplets. The inset shows the droplet size distribution (14.9 droplet/$\mu$m$^{2}$) and EDS spectrum (0.40 \% at. of Al concentration).} \label{Fig:Substrato_26nm_STT}
\end{figure}

Using different times of sputtering, Al films of 3, 10, 20, 26 and 30 nm thick were obtained. Fig. \ref{Fig:Substrato_26nm_STT} (left) shows Scanning Electron Microscopy (SEM) images of the substrate with 26 nm thick of Al where a homogeneous deposit can be observed. The inset shows a zoom in which the naked SiO$_{2}$ substrate is also visible.

After of an ex-situ annealing (see section \ref{ed}), the film turns into a homogeneous distribution of droplets, as shown in figure \ref{Fig:Substrato_26nm_STT} (right). The inset shows the histogram of the size distribution; the seed coverage has a density of 14.9 droplet/$\mu$m$^{2}$. The fit using a Gaussian function gives a mean diameter of $D_{0}$ = 22 nm and a standard deviation, $\sigma$ = 3 nm. The SEM images confirmed that the surface density of seeds increases as the thickness of the Al film was increased. 

We consider that, after deposition, the Al films were passivated by an Al oxide layer, and this could be an obstacle for the seed formation. However, the SEM images and size distribution show a very good uniformity in the size of the droplets. Therefore, we consider that the transformation of the films into particles and/or mounds by the annealing process at 600$^{\circ}$C is an evidence that, below the Al oxide layer, the Al remains in metallic form. Moreover, we can infer that the oxide layer was less than 3 nm thick, i.e., the minimum film thickness used in this work, otherwise the droplets could not have been formed.

The only elements detected by EDS on the seeded substrates, see upper right inset in Fig. \ref{Fig:Substrato_26nm_STT}, were Al, O and Si, indicating that neither the cleaning process nor the ex-situ annealing introduce detectable contamination. The concentration of Al measured was 0.40 \% at. 
 
\subsection{Effect of the Al seeds density on the growth of ZnO nanostructures.}

ZnO NSs were grown on four substrates with different catalyst densities: 3, 10, 20 and 30 nm of Al initial film thickness. The samples were named ZOAn3, ZOAn10, ZOAn20, ZOAn30, respectively. The temperature of the substrates was kept at 900$^{\circ}$C, by placing them at $D_{sf}$ = 15.0 cm from the precursor powder. A flow of Ar/O$_{2}$ mixture (20\% of O$_{2}$) was maintained and the reaction time used was 15 minutes.

\begin{figure}
\centering
\subfloat[]{
\label{Fig:ZOAn3:a} 
\includegraphics[width=11.5cm]{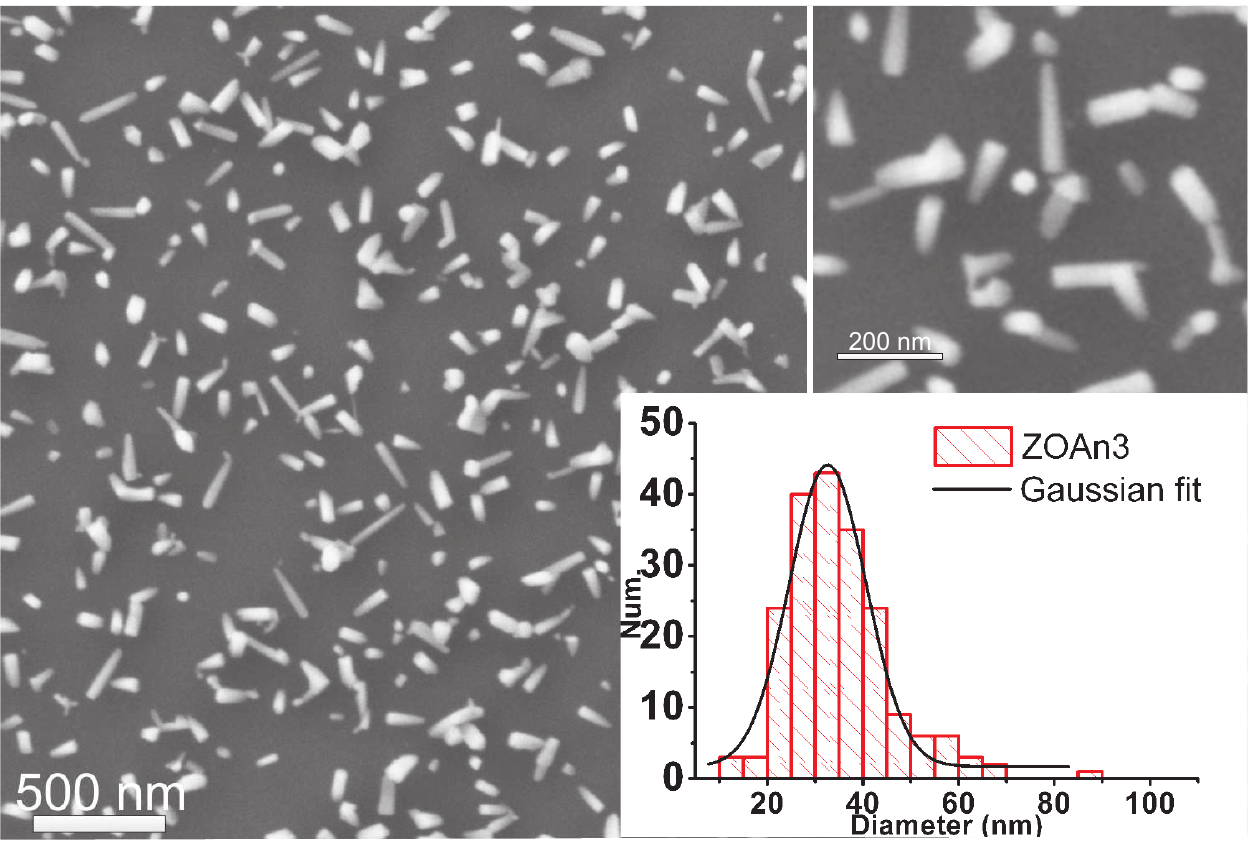}}
\hspace{0.1\linewidth}
\subfloat[]{
\label{Fig:ZOAn10:b} 
\includegraphics[width=11.5cm]{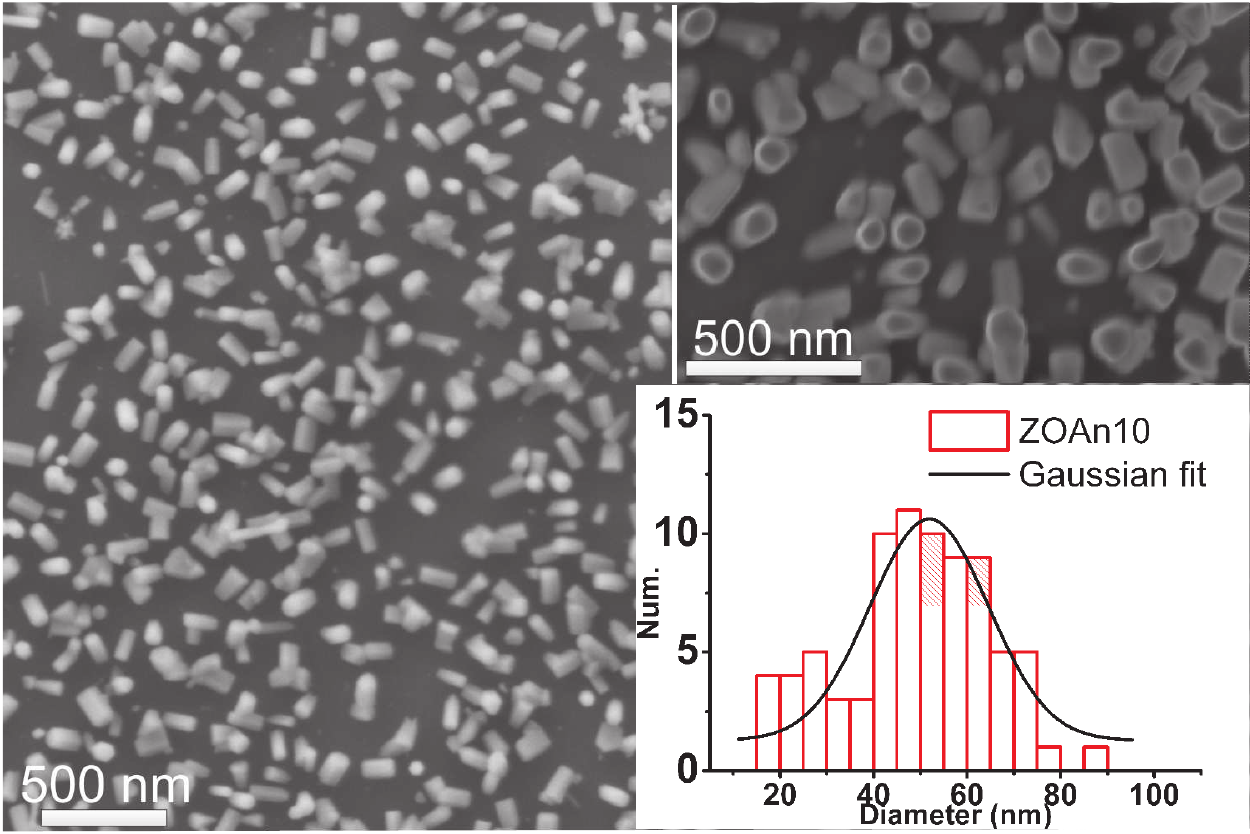} } 
\caption{a) SEM images of sample ZOAn3 and the NWs diameter distribution. On top right, a detailed SEM image of the ZnO nanowires. b) SEM images of sample ZOAn10 (detailed SEM image at top right) and the NWs size distribution with a gaussian fit.}
\end{figure}

\begin{figure}
\centering
\subfloat[]{
\label{Fig:ZOAn20:a} 
\includegraphics[width=11.5cm]{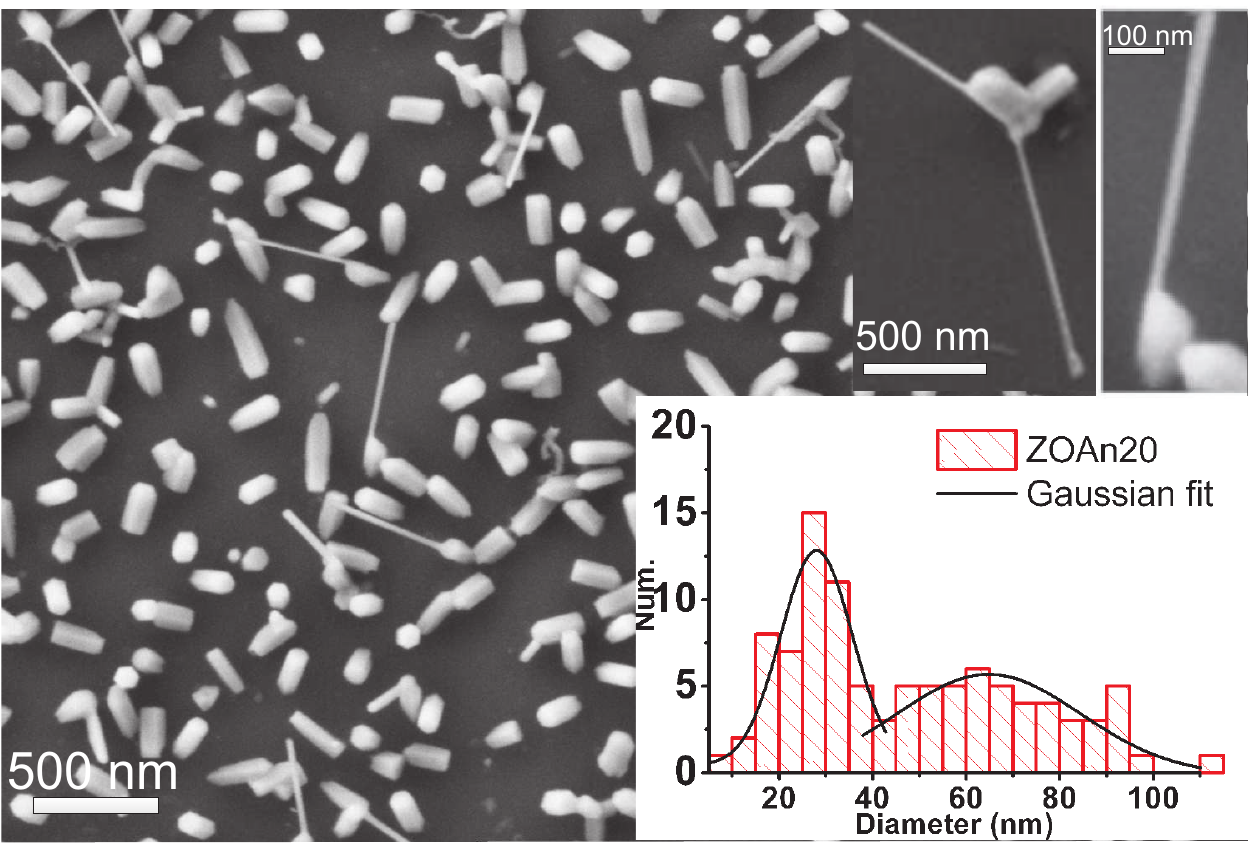}}
\hspace{0.1\linewidth}
\subfloat[]{
\label{Fig:ZOAn30:b} 
\includegraphics[width=11.5cm]{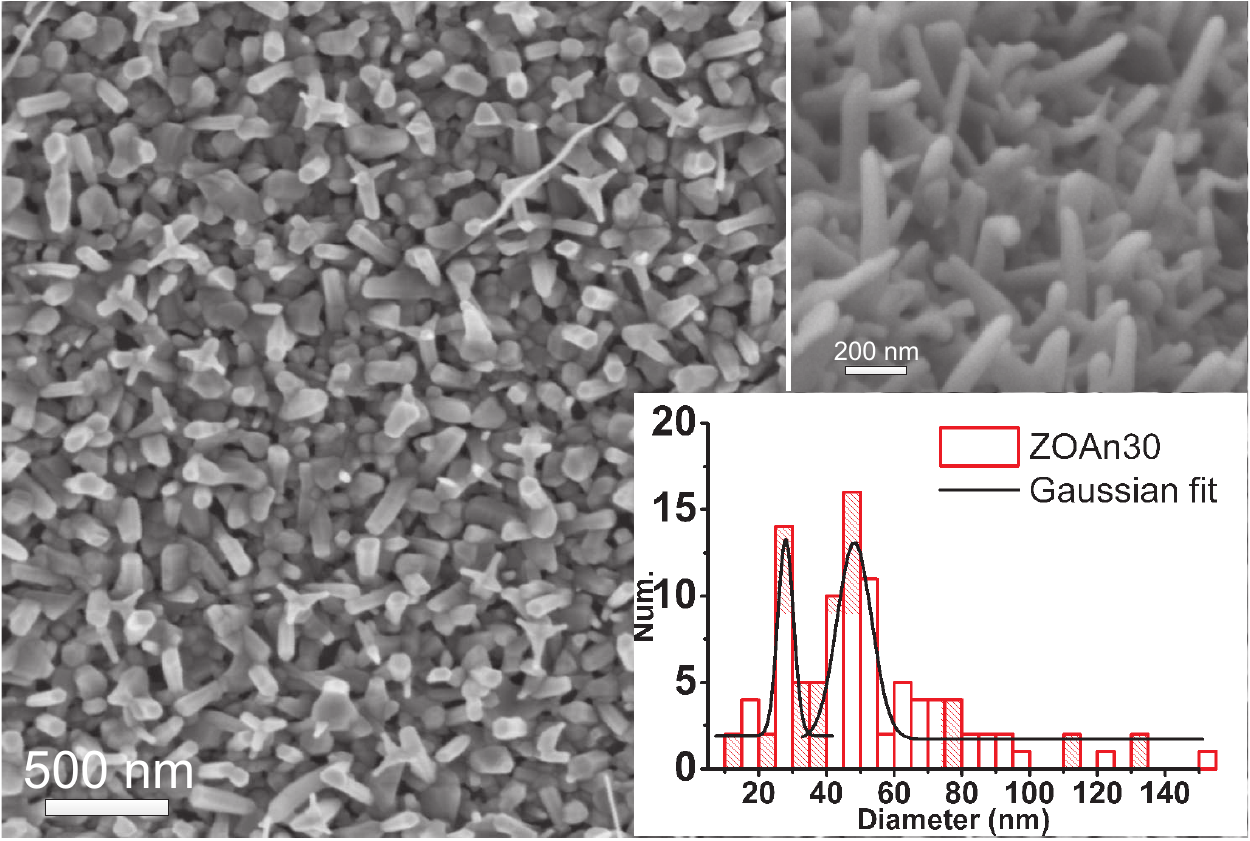} } 
\caption{a) SEM images of sample ZOAn20 (left) and the NWs size distribution (bottom right). The top right images are two enlarged views of the characteristic structures, where the faceted hexagonal shape of the nanowires and diameter reduction between growth stages can be seen. b) SEM image of sample ZOAn30. Top right shows a SEM image with a tilt of 30$^{\circ}$, and below their corresponding diameter size distribution.}
\end{figure}

Figures from \ref{Fig:ZOAn3:a} to \ref{Fig:ZOAn30:b} show the SEM images of the samples ZOAn3, ZOAn10, ZOAn20, and ZOAn30. A strong influence of the Al film thickness on the density of the grown NWs can be seen (from 24.3 to 39.7 NWs/$\mu$m$^{2}$). 

Sample ZOAn3 presents the lowest NWs coverage, see Fig. \ref{Fig:ZOAn3:a}. A mean NWs diameter: $D_{0}$ = 33 nm and a standard deviation: $\sigma$ = 8 nm were estimated from the histogram of the NWs size distribution using a gaussian fit, see at the bottom right of Fig. \ref{Fig:ZOAn3:a}. For sample ZOAn10, see Fig. \ref{Fig:ZOAn10:b}, similar structures were obtained, but with an increase in $D_{0}$ = 52 nm and $\sigma$ = 12 nm. It is remarkable that sample ZOAn20, see Fig. \ref{Fig:ZOAn20:a}, showed a clear bimodal size distribution with $D_{0}$ = 28 nm and 65 nm, and $\sigma$ = 8 nm and 15 nm, respectively.

As it can be observed in Fig. \ref{Fig:ZOAn30:b}, for sample ZOAn30, the density of the grown NWs was larger than all the previous ones. The total covering of the surface could indicate either the formation of big platelets due to the increment in the size of the Al droplets, or the formation of a ZnO film at the first stage, on which the NWs grow later. Sample ZOAn30 also presents a bimodal size distribution with $D_{0}$ = 28 nm and 48 nm, and $\sigma$ = 2 nm and 5 nm, respectively (bottom right of Fig. \ref{Fig:ZOAn30:b}).

An increase in the size of the catalyst seeds led to an increase in the size of the base of the NSs. Then, in the first stage, the ZnO grows in the form of platelets, due to an anisotropic abnormal grain growth process \cite{Tang09}. After that, the growth of thin wires followed. The dramatic diameter reduction observed (about 70\%, see Fig.\ref{Fig:ZOAn20:a}) is due to the different surface energies of the different crystal planes \cite{Tang09,Heuer78,Mitcell78}. After a certain time of growth, the structures continue to grow in the form of whiskers led by polar surface (001) of ZnO, the surface energy of this face being much higher than that of the other facets. Thus, from certain size of platelets, it is energetically favorable to change from a layer-to-layer growth mechanism to a simultaneous multilayer growth mechanism, the later being responsible of the nano-particle formation. This will be further analyzed below and in section \ref{ox}, where an increase in the time of growth was considered.\\
In order to confirm the catalyst effect of Al seeds on the ZnO NSs growth, bare SiO$_{2}$/(100)Si and Al/SiO$_{2}$/(100)Si without ex-situ annealing were used as substrate materials in a deposition process identical to that of sample ZOAn30, the one with the denser coverage, and we did not observe ZnO growth in any case.\\
These results, and the absence of NSs when no Al layer was deposited, show that the Al has indeed a catalytic behavior.

\subsection{Effect of the temperature of growth on the ZnO nanowires arrays}
\label{temp}

The substrate temperature effect was studied using Al(26nm)/SiO2/(100)Si substrates at 800$^{\circ}$C and 950$^{\circ}$C, samples ZOAtm800 and ZOAtm950, respectively. The other parameters of growth were kept without changes.

\begin{figure}
\centering
\subfloat[]{
\label{ZOAm800:a} 
\includegraphics[width=11.5cm]{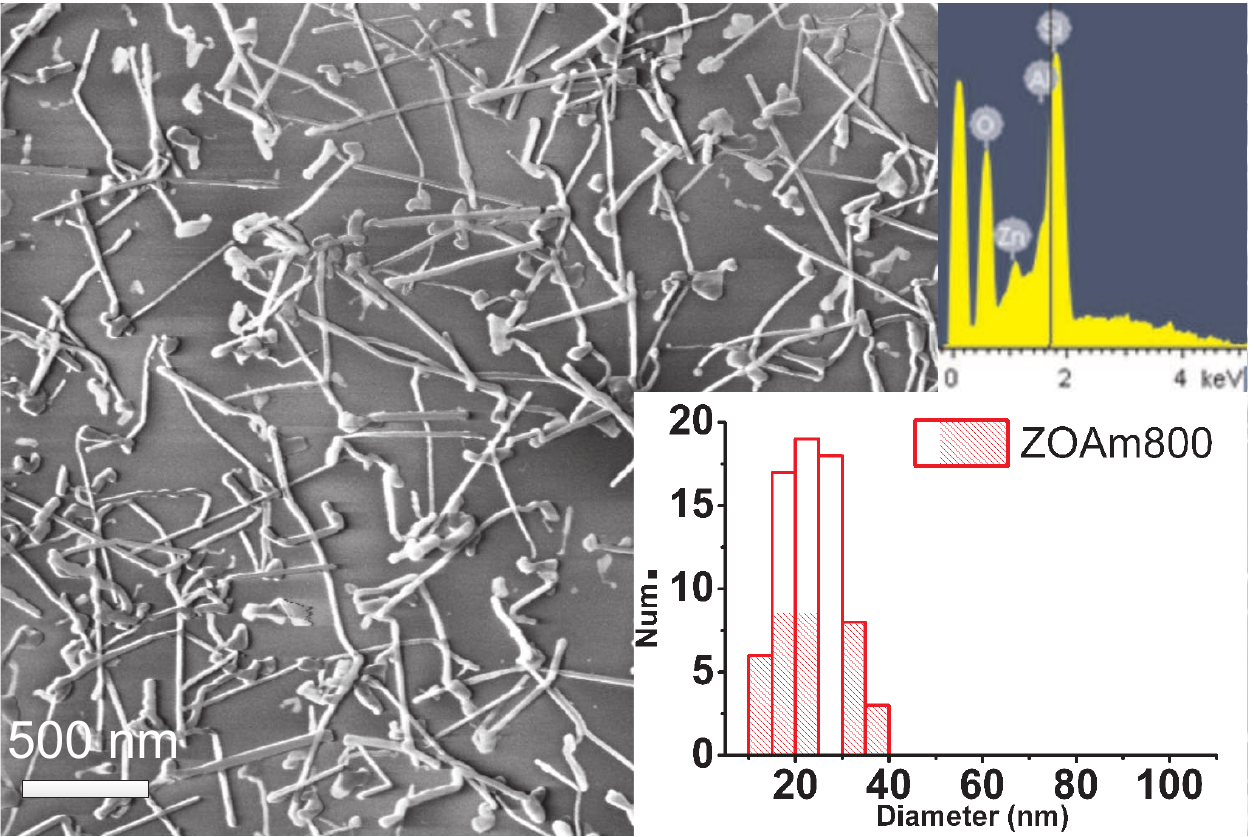}}
\hspace{0.1\linewidth}
\subfloat[]{
\label{ZOAm950:b} 
\includegraphics[width=11.5cm]{Figure5b.pdf} } 
\caption{a) SEM image, EDS spectrum and NWs size distribution of sample ZOAm800. Statistical analysis give $D_{0}$ = 23 nm and $\sigma$ = 3 nm. b) SEM image, EDS spectrum and NWs size distribution of sample ZOAm950. Statistical analysis give $D_{0}$ = 80 nm and $\sigma$ = 13 nm.}
\end{figure}

SEM images (see Fig. \ref{ZOAm800:a} and Fig. \ref{ZOAm950:b}) show a strong influence of the temperature on the morphology of the NWs. Narrow and long NWs are obtained at 800$^{\circ}$C, while the growth at 950$^{\circ}$C presents similar features than the samples grown at 900$^{\circ}$C as shown in the previous section.

The ZnO forms a dipole moment and has a polarization along the c-axis, having positively charged (Zn) and negatively charged (O) surfaces. The positive $\{$001$\}$ facet has the highest surface energy with the fastest growth rate along the c-axis, and the difference in the growth velocities of different planes/facets gives rise to various crystal facets and will determine the final shape and aspect ratio of the ZnO crystal \cite{Kong04,Soman11}.

At low temperature, the mobility and the diffusion length of the ions on the substrate are rather limited, prohibiting the ions to diffuse beyond a small range in the vicinity of the initial nucleation, and thus the unidirectional growth of narrow NWs dominates.

At higher temperatures, the rate of the Zn reoxidation is increased, and considering that the mobility and diffusion length of the ions are larger, the preferential direction of growth at the initial stage lies in the plane of the substrate.

At even higher temperatures, the mobility and diffusion length of the ions are larger, favoring platelets growth, but the rate of the Zn reoxidation is increased, thus decreasing the growth rate of the final NWs \cite{RefOxid}. The SEM image of sample ZOAm950 (see Fig. \ref{ZOAm950:b}) shows only platelets, which indicates a reduction of Zn vapor during the growth. The amount of Zn vapor that is released can be affected by a reverse oxidation reaction of Zn at temperatures higher than 900$^{\circ}$C, which is markedly dependent on the amount of CO \cite{Dadal06}. The fast oxidation of the condensing Zn gas could lead to the formation of ZnO$_{x}$ platelets, and as the ZnO has a much higher melting point (1975$^{\circ}$C) than metallic Zn, the optimum conditions required are lost, and thus no further growth would take place.

Figs. \ref{ZOAm800:a} and \ref{ZOAm950:b} also show the EDS spectra of the samples grown at 800 and 950$^{\circ}$C, from which Zn concentrations of 1.03 and 0.35 \%at., respectively can be determined. The value of Al concentration is in total agreement with the expected, about 0.40 \% at. in both samples, since they had the same thickness of initial Al film. No trace of other metal components can be observed, corroborating that the experimental procedure is clean and free of detectable contamination.

\subsection{Influence of the partial pressure of oxygen during the growth.}
\label{ox}

The total pressure and oxygen partial pressure play an important role in the kinetics of the ZnO condensation and the control of the characteristics of the NWs \cite{Dadal06}. While the Ar flow is associated only to the gas transport, the oxygen partial pressure is often associated to the growth process in ZnO carbothermal reduction \cite{Klaus11}. Based on this, different oxygen concentrations were injected in the gas flow during the growth of the NWs, preparing the samples ZOAp10, ZOAp20, and ZOAp30, where the number stands for the percentage of oxygen concentration. The samples were grown on Al(26nm)/SiO2/(100) Si substrates at 900$^{\circ}$C using a longer deposition time of 30 min., in order to enhance the influence of the oxygen concentration variation.

\begin{figure}
\includegraphics[width=1\columnwidth]{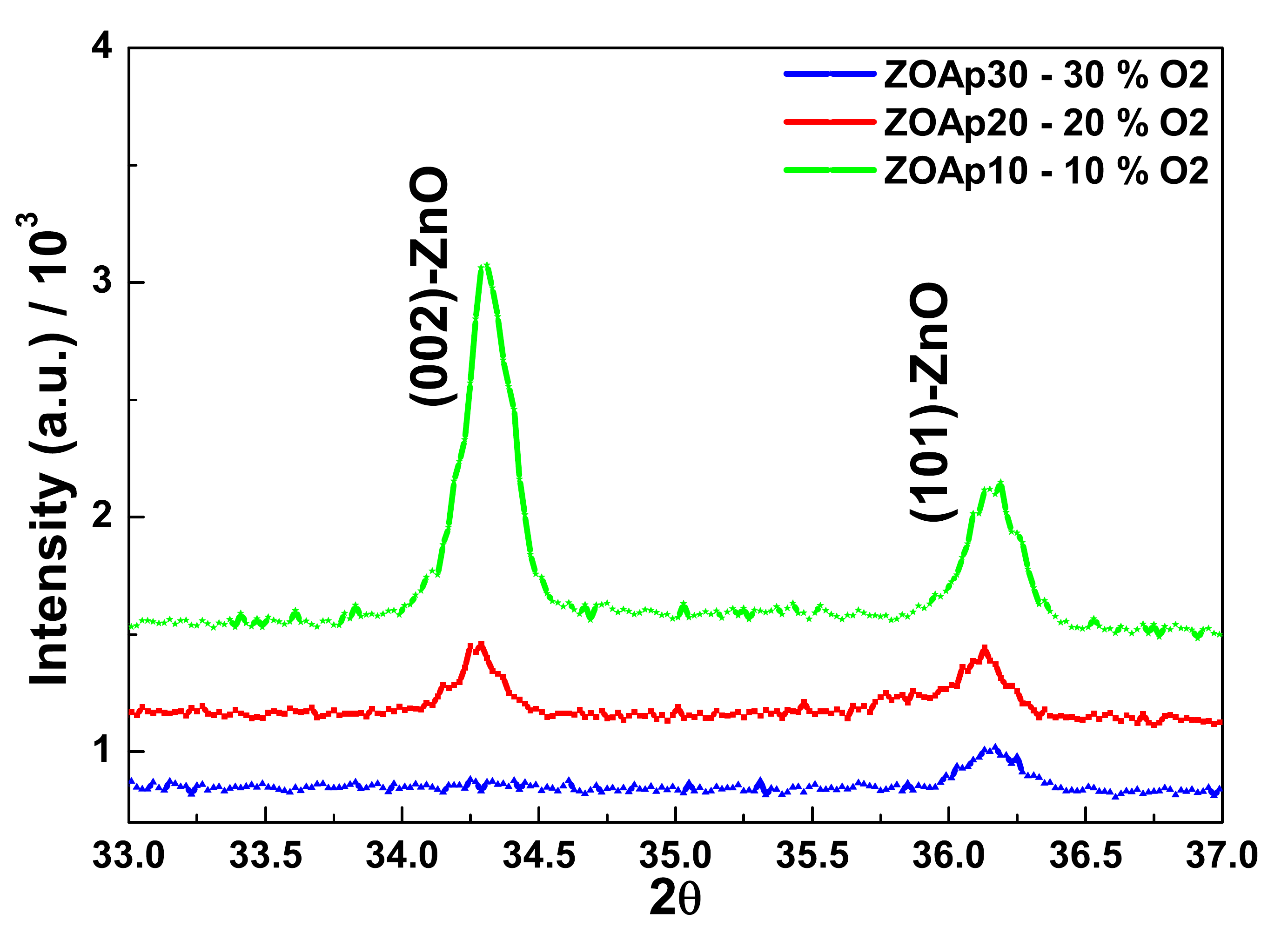}
\caption{X-ray diffraction spectra of the samples grown with 10, 20 and 30 percent oxygen of the 150 sccm Ar/O$_{2}$ mixture.} \label{XRD_oxig}
\end{figure}
\begin{figure}
\includegraphics[width=1\columnwidth]{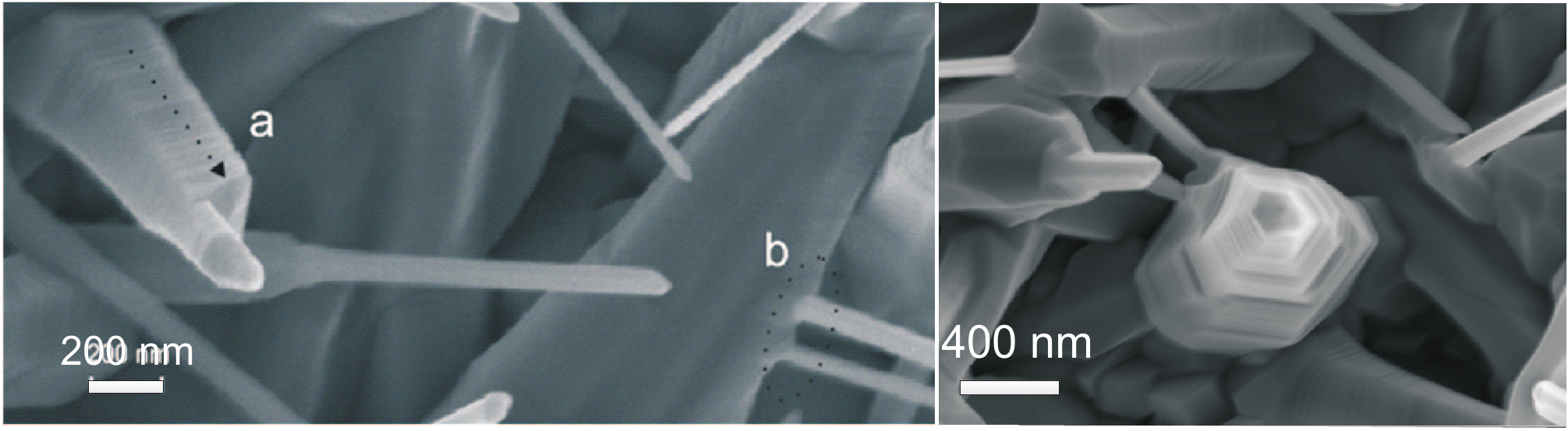}
\caption{SEM image of sample ZOAp10. (Left) The growth direction marked with an arrow, (a), and a possible self-catalysis process, (b), are shown. (Right) Hexagon-shaped pyramids with flat terraces and steps are seen at the ends of the NWs.} \label{Fig:Detall_Ox}
\end{figure}
\begin{figure}
\centering
\includegraphics[width=1\columnwidth]{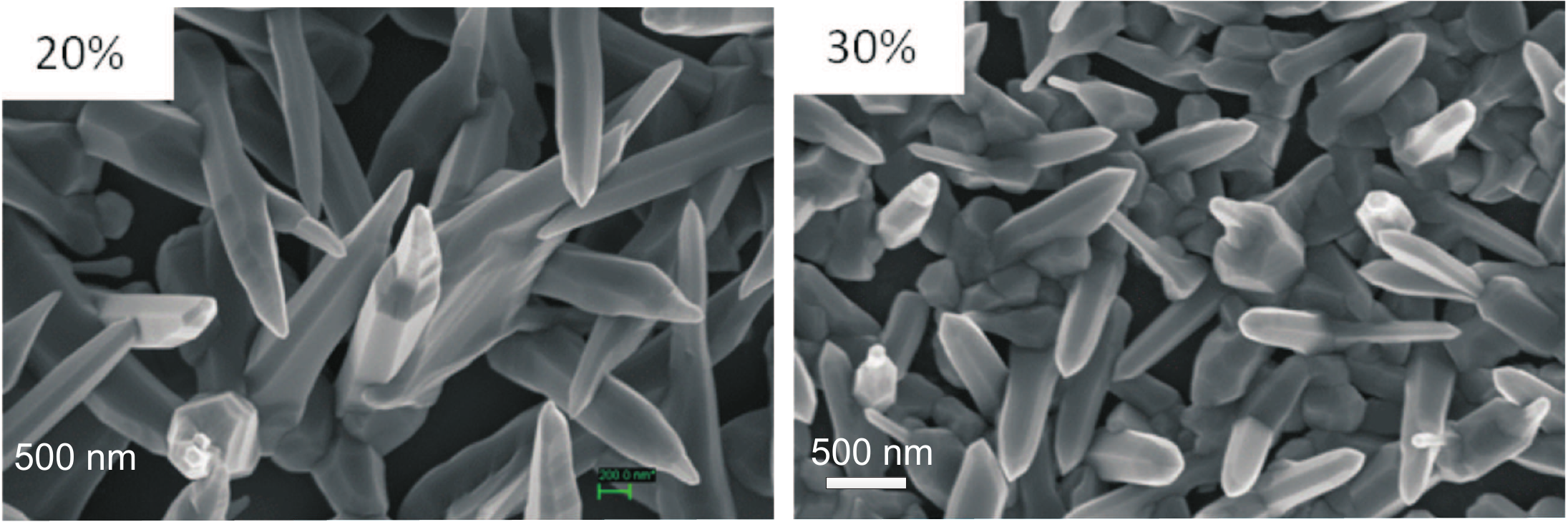}
\caption{SEM images of the sample ZOAp20 (left) and ZOAp30 (right). The increase of the oxygen partial pressure  gives rise to a higher covering and homogeneity of the NWs. In both cases, the tips are strongly faceted or flat.}
\label{Fig:Ap20y30}
\end{figure}
\begin{figure}
\centering
\includegraphics[width=1\columnwidth]{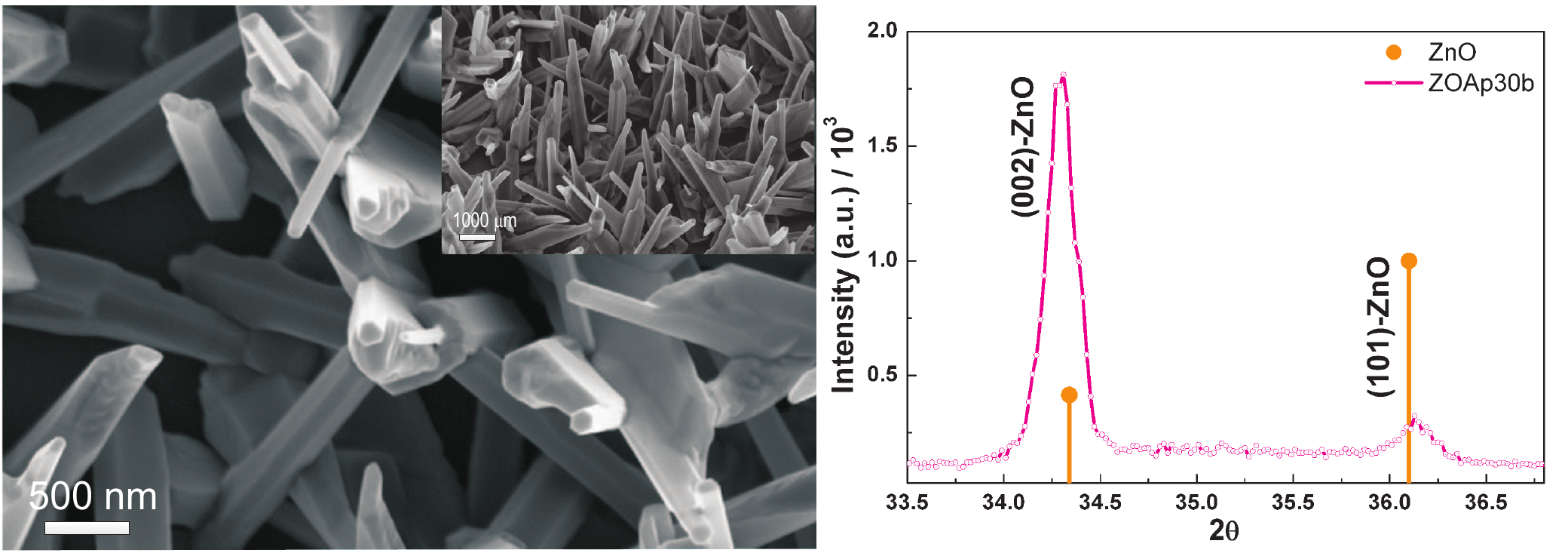}
\caption{SEM image of sample ZOAp30b growth at 800$^{\circ}$C and its corresponding diffractogram (ZnO:chart79-208 \cite{RefPCPDFZnO}). In the inset, the same sample with a tilt of 30$^{\circ}$ is shown.}
\label{Fig:ZOAp30_800}
\end{figure}

Fig. \ref{XRD_oxig} shows the X-ray diffractograms of these three samples, where a strong influence of the oxygen pressure in the preferential orientation of the crystal structure of the NWs can be seen. For the three oxygen concentrations studied, nanowire-like structures were observed (see Figs. \ref{Fig:Detall_Ox}, \ref{Fig:Ap20y30}, \ref{Fig:ZOAp30_800}). 

In bulk polycrystalline ZnO, the intensity corresponding to (002) plane is about 40 $\%$ lower than the intensity of the main (101) reflection \cite{RefPCPDFZnO}. In sample ZOAp10, the intensity corresponding to the (002) plane is about 300$\%$ higher than the intensity of the (101) reflection indicating that there is a marked texture in the (001) direction, normal to the substrate surface. An increase of the oxygen percentage causes a reduction in the preferential orientation: similar intensities of the (002) and (101) diffraction peaks are observed with 20\% O$_{2}$, and a totally polycrystalline structure occurs for 30\% of oxygen during the growth. The c-lattice parameter of ZnO as calculated from the positions of the (002) reflections is c=5.23 \AA, which correlates well with the c value of 5.2069 \AA \hspace{0.2cm} for bulk ZnO \cite{RefPCPDFZnO}, indicating that a low strain level exists in the NWs without significant changes regarding the partial pressure of oxygen. 


The three samples show the same growth mode, but an increment in the covering is observed as the oxygen concentration is increased. Fig. \ref{Fig:Detall_Ox} shows in detail the different stages of growth for the sample with low oxygen partial pressure. At the left, the direction of growth of the consecutive hexagon-shaped pyramids with flat terraces is marked with an arrow and marked (a). In the same figure, an example of growth on the sides of the main NWs, marked (b), is shown. These lateral growths are strongly reduced (in quantity) when the partial pressure of oxygen is increased to 30 $\%$, what can be associated with Zn-rich regions (growth lead by self-catalysis from ZnO droplets attached to the surface and tip of the rods).

A fourth sample, ZAOp30b, was prepared using a lower growth temperature of 800$^{\circ}$C and maintaining the rest of the parameters equal to those of sample ZOAp30. The typical growth mode observed in the previous samples is obtained (Fig. \ref{Fig:ZOAp30_800} (left)). But, in contrast with what was found in sample ZOAp30, sample ZOAp30b presents a preferential growth in the (00l) plane (Fig. \ref{Fig:ZOAp30_800} (right)). This indicates that a reduction in the temperature causes similar consequences than those of a reduction in the oxygen partial pressure during the growth, at least with respect to the crystallographic orientation.

\subsection{Growth mechanism}
\label{resGrowth}

\begin{figure}
\includegraphics[width=1\columnwidth]{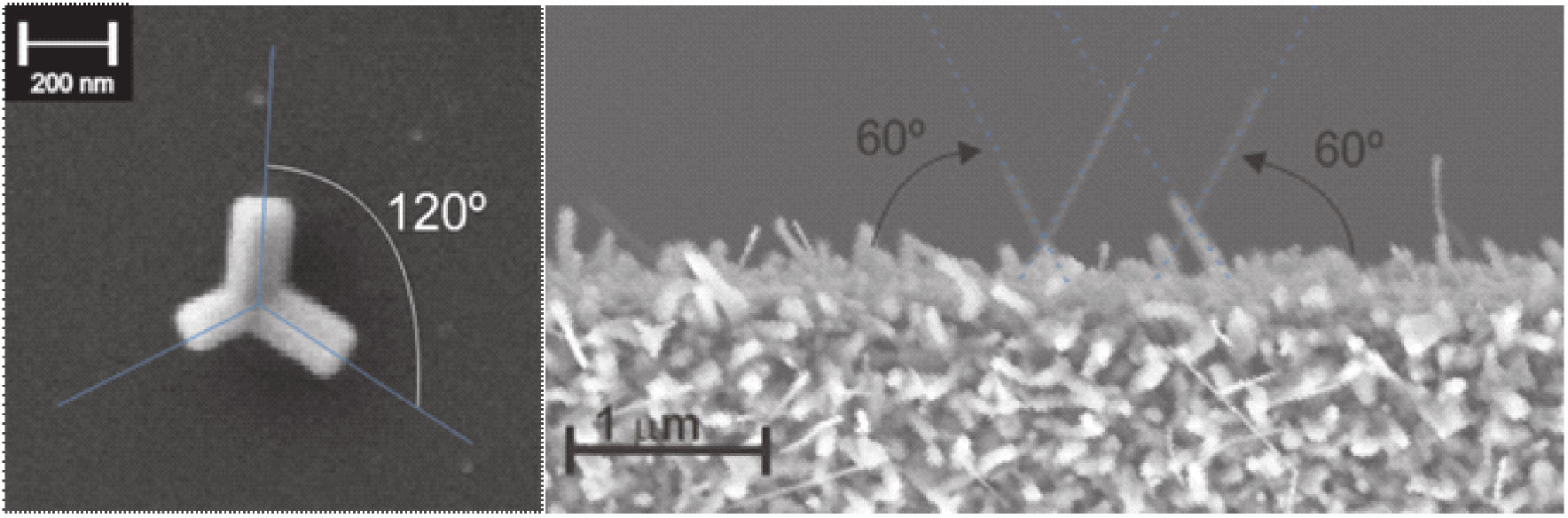}
\caption{(Left) A twin stacking fault gives rise to an azimuth of 120$^{\circ}$ between NWs (SEM image taken from sample ZOAn30). (Right) Side view of the sample. The indicated angle is measured with respect to the substrate surface.} \label{Fig:AngulosNWs}
\end{figure}

So far, the effect of the Al as catalyst is evident, it remains to be determined whether the NWs growth occurs via VLS, VS, or VSS process. 
For a study of the growth modes, SEM images were analysed considering the directions of NWs, as is shown in Fig. \ref{Fig:AngulosNWs}. Tripod-like structures are observed to appear with increasing thickness of Al initial films. The 120$^{\circ}$ angle between NWs in this stacking fault could be indicating that the growth occurs starting from pyramidal mounds with $\{$11$\bar{2}$3$\}$ planes as faces. \cite{Baxter03}. These structures are frequently observed in the self-catalysis process \cite{Fortuna10,Mattila07}.

It is well known that the aluminium surface oxidation tends to be self-limiting, and beyond a surface layer a few nanometres thick, there is no further oxidation \cite{Campbell99,Reichel08}. Besides, H. Oughaddou \textit{et al. }\cite{Oughaddou05} showed that the oxidation process (at room temperature) stabilizes the aluminum atoms at the surface by forming a stable compound (AlO$_{x}$), rather than bulk alumina (Al$_{2}$O$_{3}$). We consider that in our experiments this AlO$_{x}$ compound could have formed a diffusion barrier for the Zn atoms, thus promoting the nucleation and growth of Zn in the initial stage. It is worth to remark, that in all the growth processes of our samples, the same heating rate (25$^{\circ}$C/min) was used, so we consider that the aluminium surface oxidation process, previous to the nucleation and growth process of ZnO, was the same in all the samples. Furthermore, no Al$_{2}$O$_{3}$ phase was detected by XRD in any of our samples. The conventional VLS process can be discarded as the actual growth mechanism, since an eutectic mixture of Al-Zn cannot be obtained at a temperature of 900 $^{\circ}$C \cite{Zhu04}. Moreover, for this process, the diameter, length and density of the wires are primarily defined by the size and density of the catalyst droplets, whilst in our results, the initial thickness of the catalyst film has no influence on the diameter of the NWs. Also, these NWs present a cylindrical-faceted shape, and have a tendency to be tapered, which is a typical characteristic of VSS growth \cite{YWang06,Potie11}. Unlike the typical Au-catalised NWs, our NWs do not have a spherical cap at the top. Instead, the tips result strongly faceted or flat. This fact suggests that the catalyst does not melt, otherwise surface tension forces would favor a spherical shape of the tips \cite{Potie11}. However, it is likely that in our case the nucleation and initial growth of NWs occur in two stages, namely an initial process of diffusion of Zn in Al$_{2}$O$_{3}$/AlO$_{x}$ with subsequent Zn oxidation, followed by a process where the ZnO$_{x}$ acts as self-catalyst that leads to the formation of the NSs.

In the first stages of the growth, AlO$_{x}$/Al islands are the preferential sites for Zn incorporation since the SiO$_{2}$ surface has a relatively lower reactivity. This mechanism is evidenced by the absence of material deposited on the bare substrate (without Al-catalyst). 

VS mode is another way of getting the nucleation and later growth of the NWs. It is possible that either atomic steps on the substrate formed by screw dislocations, or ditches at the junction of the two grain surfaces of twinned structures, act as sites where a continuous deposition occurs, but the morphology of the resulting NSs is totally different from those observed in our samples \cite{Klaus11}. This is the reason why VS is discarded as the actual growth mode, at least as initial growth mechanism.

In the light of these considerations, the main growth mode appears to be occurring through a VSS process.


The NWs growth occurs by the piling up of consecutive hexagon-shaped layers of similar size, but an abrupt change in the size of the resulting islands occurs at some moment, see for example, the region marked as (a) in the Fig. \ref{Fig:Detall_Ox}(left). As it was mentioned previously, the growth mode can be explained considering that after the Al-assisted nucleation and subsequent whisker growth, there is a change in the mode from a layer-by-layer growth to a simultaneous multilayer growth. H. Tang \textit{et al.} \cite{Tang09} reported in their detailed study of the growth mechanism of ZnO NWs that this abrupt change in the diameter can be reasonably explained by the ES barriers model \cite{Ehrlich66,Schwoebel69}. Then, under certain conditions an ES activation barrier could be ruling the interlayer mass transport at the edge of the (001) plane and this could lead to the aforementioned mode change. If the ES barrier effect is strong, those atoms deposited on top of monolayer islands are inhibited from diffusing over the edges, resulting in an increased nucleation rate of the new layer.

Alternatively, another interpretation for NWs nucleation was proposed by G. Perillat-Merceroz \textit{et al.}
\cite{Perillat2013}, where they claim that the crystal polarity has a crucial role in the nucleation, growth and shape of the ZnO nanostructures. They showed that using substrates with defined polarities, it is possible to control the growth of the nanostructures, their samples having similar characteristics to those of ours examined in Fig. \ref{Fig:ZOAp30_800}. There, the "pyramids" and nanowires on the top of the pyramids have opposite polarities, being the NWs Zn-polar and the pyramids O-polar. This inversion of the domain boundaries could be induced by Al impurities that in our case can diffuse from the catalyst droplet. This can be an alternative explanation to the lateral growth observed in the NWs shown in Fig. \ref{Fig:Detall_Ox}.

\section{Conclusions}

We report the successful growth of Al-catalysed ZnO NWs. The NWs were grown on SiO$_{2}$/Si substrates covered with different Al droplet sizes and densities. Our results indicate that the Al effectively acts as a catalyst, not ruling the diameter of the NSs as in the VLS processes, but controlling other aspects of their morphology during the growth. The growth temperature was found to be a critical parameter in the set of better conditions for the initial nucleation. The partial pressure of oxygen was shown to have a strong influence in the structural properties of the NWs, thus becoming a parameter through which the preferential crystallographic orientation can be controlled. The different studies in this paper show that neither VLS nor VS are the main mechanisms of growth when Al is used as catalyst. Instead, we reasonably conjecture that a VSS-like mechanism is mainly present. Nevertheless, more experiments are needed to deepen on the details of the mechanism of growth.\\ 
Our results indicate that Al works as an excellent catalyst for the production of ZnO NWs and thus offers promising perspective for the industry, since Al costs are significantly lower than those of the conventional catalysts.\\

\section*{Acknowledgements}
This work was partially supported by FONARSEC-FSTICs 002/2010, CIUNT Prog. 26/E439 and ANPCyT-PICT-2010-0400. We thank C. Zapata for her technical assistance. We thank H. Brizuela for careful reading of this manuscript.

\section*{References}

\bibliographystyle{unsrt}

\end{document}